# Predicting nonlinear dynamics of optical solitons in optical fiber via the SCPINN

Yin Fang, Wen-Bo Bo, Ru-Ru Wang, Yue-Yue Wang [*] and Chao-Qing Dai [*]

*College of Optical, Mechanical and Electrical Engineering, Zhejiang A&F University, Lin'an, Zhejiang 311300, P. R. China*

**Abstract.** The strongly-constrained physics-informed neural network (SCPINN) is proposed by adding the information of compound derivative embedded into the soft-constraint of physics-informed neural network(PINN). It is used to predict nonlinear dynamics and the formation process of bright and dark picosecond optical solitons, and femtosecond soliton molecule in the single-mode fiber, and reveal the variation of physical quantities including the energy, amplitude, spectrum and phase of pulses during the soliton transmission. The adaptive weight is introduced to accelerate the convergence of loss function in this new neural network. Compared with the PINN, the accuracy of SCPINN in predicting soliton dynamics is improved by 5-11 times. Therefore, the SCPINN is a forward-looking method to study the modeling and analysis of soliton dynamics in the fiber.

## 1. Introduction

When the optical pulse propagates along the fiber, the waveform, amplitude and speed of pulse remain unchanged. Such pulse is used in all-optical soliton communication to overcome the limitations of speed and distance in the linear wave transmission system [1,2]. In the single-mode fiber, the picosecond optical pulse happens the symmetrical broadening of pulse time and frequency domains respectively due to the effects of group-velocity dispersion (GVD) and the self-phase modulation (SPM). When these two effects achieve a complete mutual balance, bright and dark solitons form in the abnormal and normal dispersion regions respectively [3,4]. When the pulse width reaches the order of femtosecond or subfemtosecond, the influence of various higher-order effects on the pulse can not be ignored [5,6]. Among these effects, the third-order dispersion (TOD) has the least influence on the fundamental soliton, and thus it can be regarded as the perturbation, while it makes the higher-order soliton produce splitting phenomenon. The self-frequency shift (SFS) effect will cause the pulse spectrum to produce a red shift phenomenon. The self-steepening (SS) effect will lead to the displacement of chirped pulse in time and frequency domains.

The transmission dynamics of optical pulse in optical fiber is governed by the nonlinear

---

[*] Corresponding author email：wangyy424@163.com (Y.Y. Wang); dcq424@126.com (C.Q. Dai)

Schrödinger equation (NLSE) [7, 8]. In the past, analytical methods [9-11] were mainly used to obtain accurate solutions. Although these methods are effective for standard NLSE, with the deepening of research, especially when all order dispersion, nonlinearity and various effects coexist, it is usually difficult to find the analytical solution, let alone specific form of analytical solutions. Therefore, researchers focused on numerical methods such as the finite-difference [12], pseudo spectral method [13] and finite elements [14] to obtain approximate solutions. Although great progresses have been made in solving numerical solutions, the parameter selection techniques of these numerical methods are relatively complex to implement and they are extremely unfriendly to researchers without numerical experience. Moreover, to obtain high-precision approximate solutions, the numerical method needs to generate a large number of grid points in the time domain, which is a very time-consuming process. And it is difficult to infer the control parameters of the physical model from the limited data. Therefore, it is necessary to introduce more advanced methods to replace traditional numerical simulation and realize the approximate solution of NLSE.

The deep learning and artificial intelligence have always been regarded as one of the most breakthrough technologies in this era [15,16]. Traditional deep learning is a function approximator developed based on pure data, which is difficult to analyze the internal input-output relationship of models.This function approximator can reduce the computational cost when dealing with nonlinear problems, and it is easy to be popularized. However, its dependence on data quality and size has become the main reason to limit its wide application. As a meshless method, the PINN alleviates [17] the problem of dimension explosion of traditional numerical and data dependence of traditional deep learning to some extent, and is widely used in many partial differential equations (PDEs). To solve more complex physical models or complex solution areas, the PINN algorithm has also been improved, including CAN-PINN [18], VPINNs [19], hp-VPINNs [20] and so on. Recently, the convergence [21] of the PINN has been proved theoretically.

In the field of optical solitons, Chen et al. applied the multilayer PINNs to study the data-driven rogue wave, breather wave and periodic wave solutions of Chen-Lee-Liu equation [22]. Su et al. revealed the wave-packet evolution behavior in the optical fiber via using the modified PINN [23]. Yan et al. studied the forward and inverse problems of the logarithmic NLSE with PT-symmetric harmonic potential [24] and the forward and inverse problems of Hirota equation [25] by the PINN. We mainly pay attention to integrable equations that describe the pulse evolution in optical fiber. These PDEs have rich exact solutions of optical solitons, which provide many samples for the PINN training. Meanwhile, these kind of systems has many excellent physical properties, such as the conservation law, the symmetry and the Lax pair, which provide many ideas for the improvement of the PINN. Currently, aiming at the standard NLSE [26,27], Manakov equation [28] and higher-order nonlinear Schrödinger equation (HNLSE) [29,30], we have reported the dynamic behaviors of one-soliton, two-soliton and rogue wave in the single-mode fiber via the PINN and its improved method. However, the above studies [22-29] have neglected the influence of physical effects on soliton transmission. In addition, for ensuring the accuracy of prediction, most of the above improved methods [18-20] only focus on a small range of solution areas.

For fully understand the formation of optical solitons and realize the pulse prediction of longer transmission distance in optical fiber. We design the PINN structure with the parallel two-subnets, so as to embed the physical information of composite derivative into the PINN, and construct a PINN method with stronger soft-constraint, namely the SCPINN. Applying this method to analyze the influence of various physical effects on optical solitons transmission in single-mode fiber. Simultaneously, more adaptive variables such as adaptive weight and adaptive activation function [31] are introduced to accelerate the convergence of loss function. This paper has the following novelty. (I). The formation process of bright and dark solitons, and two-soliton molecule in optical fiber is predicted by a new neural network. (II). The changes of physical quantities such as power conservation, phase, spectrum and amplitude are analyzed. (III). Compared with the classical PINN, the accuracy of soliton pulse evolution predicted by this new neural network is improved about one order of magnitude.

## 2. Physical model and SCPINN method

The transmission of femtosecond optical pulses in single-mode fiber can be described by HNLSE, which also characterizes various physical effects in single-mode fiber [32]. Describing the impact of these physical effects on soliton transmission can deepen our understanding of the formation of solitons in optical fiber and provide theoretical guidance for realizing high-capacity, long-distance femtosecond soliton communication, and it has the form

$$iQ_z + \lambda_1 Q_{tt} + \lambda_2 |Q|^2 Q + i[\lambda_3 Q_{ttt} + \lambda_4 (|Q|^2 Q)_t + \lambda_5 Q(|Q|^2)_t] = 0, \tag{1}$$

where $Q, z, t$ are the complex envelope of the slow-varying optical field, propagation distance and delay time normalized by $\sqrt{p_0}, L_D, T_0$ respectively with the peak power $p_0$, the half width $T_0$ of the incident pulse and the GVD length $L_D = T_0^2 / |\beta_2|$. The normalized quantities $\lambda_1, \lambda_2, \lambda_3, \lambda_4, \lambda_5$ reflect the intensity of effects of GVD, SPM, TOD, SS, and SFS, which are equal to 1/2, $L_D / L_{NL}$, $L_D / (6L_D^{'})$, $1/(\omega_0 T_0)$, $T_R / T_0$ respectively with the TOD length $L_D^{'} = T_0^3 / |\beta_3|$ and the nonlinear length $L_{NL} = 1/(\gamma p_0)$. $\beta_2, \beta_3, \gamma$ are coefficients of GVD, TOD and nonlinear respectively, $\omega_0$ is the carrier frequency and $T_R$ is the Raman resonant time constant. Considering the Ytterbium-doped fiber with the central wavelength $\lambda_0 = 1060 nm$ [33], the reference values are $\beta_2 = 23 fs^2 / mm$, $\beta_3 = 63.8 fs^2 / mm$, and $\gamma = 4.7 \times 10^{-6} w^{-1} / mm^{-1}$. When the physical parameters take $\lambda_1 = 0.5, \lambda_2 = 1, \lambda_3 = \lambda_4 = \lambda_5 = 0$, Eq. (1) degenerates into the focusing NLSE [34], where the GVD and SPM interact to form bright soliton in the anomalous dispersion region of the fiber. Taking the parameter $\lambda_1 = 0.5, \lambda_2 = -1, \lambda_3 = \lambda_4 = \lambda_5 = 0$, Eq. (1) becomes a defocusing NLSE that can stably exist dark soliton [35].

To accurately predict the influence of various physical effects on optical pulses. As shown in Fig. 1, we consider the feedforward neural network structure of the parallel two-subnets. The design of the parallel two-subnets introduces more loss terms of the loss function. Experience and theory [36, 37] show that the increase of loss term improves the robustness and generalization ability of the network, which makes the prediction more accurate. The feedforward network is relatively simple, but it can solve most of the problems of PDEs. Subnet-1 learns the solutions $r, m$ of the governing equation, and subnet-2 learns the compound derivative $r_t, m_t$ or $r_{tt}, m_{tt}$. Via the pseudo spectral method and Fourier derivative property, we obtain the data set of solution $Q$, compound derivative $Q_t$ or $Q_{tt}$, all of which are (512×401). The prediction results of adding the first-order compound derivative are more accurate than the 2nd-order compound derivative (See section 4 discussion for details). Therefore, the SCPINN with the first-order compound derivative is used in all cases in this paper. $Q = r + im$ is the dimensionless function with its real and imaginary parts $r, m$, so $|Q|^2 = QQ^*$ in Eq. (1), $Q^*$ is the conjugate of $Q$. This model is derived from the classical NLSE by using the multi-scale method [32], and it is mainly used to describe the real physical situation of femtosecond optical pulse transmission in optical fibers. The real and imaginary parts of Eq. (1) are separated as

$$
\begin{aligned}
f_r &= m_z - \lambda_1 r_{tt} - \lambda_2 (r^2 + m^2) r + \lambda_3 m_{ttt} + \lambda_4 ((r^2 + m^2) r)_t + \lambda_5 (r^2 + m^2)_t r, \\
f_m &= i(r_z + \lambda_1 m_{tt} + \lambda_2 (r^2 + m^2) m + \lambda_3 r_{ttt} + \lambda_4 ((r^2 + m^2) m)_t + \lambda_5 m (r^2 + m^2)_t).
\end{aligned}
\tag{2}
$$

Considering the initial / boundary conditions of the governing equation and the first-order compound derivative, the loss function of the neural network with the adaptive weight is

$$
Loss = \kappa_a \Gamma_{IC} + \kappa_b \Gamma_{BC} + \kappa_c \Gamma_{PDE} + \kappa_d \Gamma_{DO}, \tag{3}
$$

in which

$$
\begin{aligned}
\Gamma_{IC} &= \frac{1}{N_0} \sum_{\eta=1}^{N_0} (|r(0, t^\eta) - r^\eta|^2 + |m(0, t^\eta) - m^\eta|^2 + |r_t(0, t^\eta) - r_t^\eta|^2 + |m_t(0, t^\eta) - m_t^\eta|^2), \\
\Gamma_{BC} &= \frac{1}{N_b} \sum_{q=1}^{N_b} (|r(z^q, t^q) - r^q|^2 + |m(z^q, t^q) - m^q|^2 + |r_t(z^q, t^q) - r_t^q|^2 + |m_t(z^q, t^q) - m_t^q|^2), \\
\Gamma_{PDE} &= \frac{1}{N_f} \sum_{j=1}^{N_f} (|f_r(z^j, t^j)|^2 + |f_m(z^j, t^j)|^2).
\end{aligned}
\tag{4}
$$

And $\kappa_a, \kappa_b, \kappa_c, \kappa_d$ are the automatically updated weight variables introduced into the neural network. Their initial values are set to 1. Here, the derivative of the function in *Loss* is obtained by the automatic differentiation, which has been implemented in the tensorflow software [38] package. $\Gamma_{IC}, \Gamma_{BC}$ represent the mean square error (MSE) between the output of 1/2 of the subnet and the initial / boundary conditions of $Q, Q_t$, and $\Gamma_{PDE}$ is the MSE of the governing equation at the configuration points. $\Gamma_{DO}$ represents the MSE at the configuration points between the output

$r_t^{AD}, m_t^{AD}$ of the automatic differentiation of subnet-1 and the output $r_t, m_t$ of subnet-2, which is also the constraint relationship between the subnet-1 and subnet-2. Its form is

$$\Gamma_{DO} = \frac{1}{N_f} \sum_{p=1}^{N_f} \left( \left| r_t(z^p, t^p) - r_t^{AD}(z^p, t^p) \right|^2 + \left| m_t(z^p, t^p) - m_t^{AD}(z^p, t^p) \right|^2 \right). \quad (5)$$

The parameters $NN(w, b, a, \kappa_v)$ in the network are optimized by minimizing the above loss function, where $w, b$ represent the weight and bias of the feedforward neural network respectively, and $a$ is the slope of the adaptive activation function. In this paper, we uniformly adopt the initial / boundary sampling points $N_0 = N_b = 100$, and the configuration point $N_f = 20000$, these sampling points are random sampling without repetition. In each subnet, the network structure is 5 layers*30 neurons. Moreover, we use the Adam optimizer to optimize the loss function (3)

For quantitatively describe the predicted accuracy of the SCPINN, we introduce the relative error $L_2$ as a reference quantity, and its expression is

$$L_2 = \frac{\sqrt{\sum_{1 \leq \delta \leq 401, \ 1 \leq \vartheta \leq 512} \left[ Q^{pred}(z^\delta, t^\vartheta) - Q(z^\delta, t^\vartheta) \right]^2}}{\sqrt{\sum_{1 \leq \delta \leq 401, \ 1 \leq \vartheta \leq 512} Q^2(z^\delta, t^\vartheta)}}. \quad (6)$$

The relative error $L_2$ is calculated by all reference points $(512 \times 401)$ in the data set.

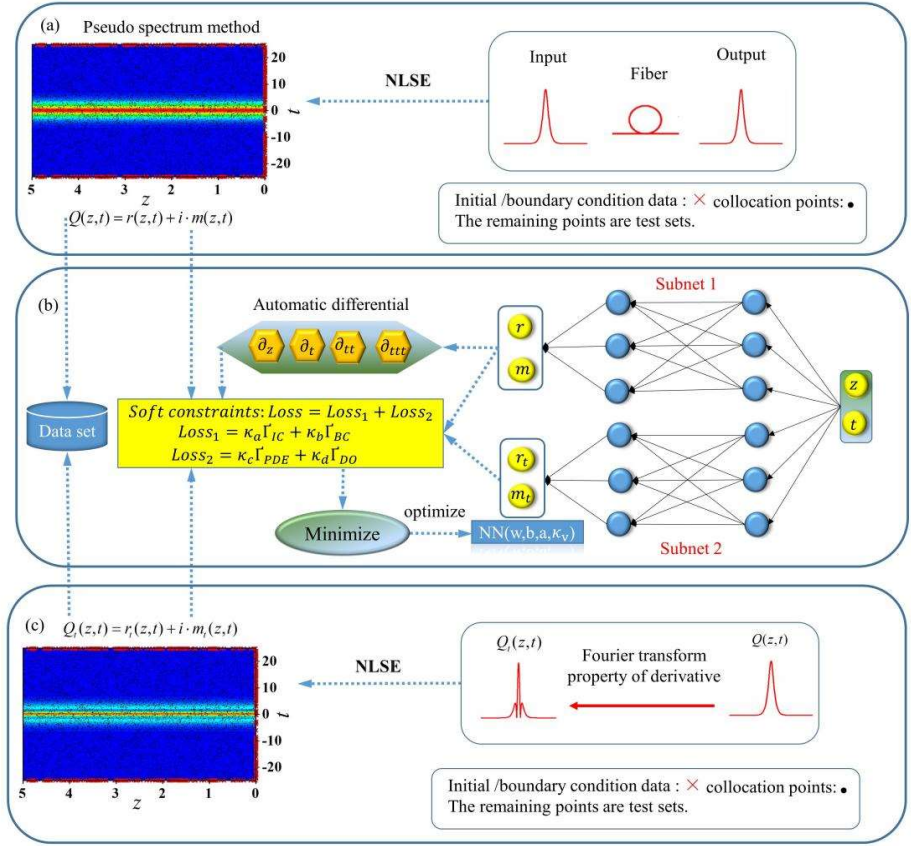

**Fig.1.** The SCPINN predicts the nonlinear dynamics of optical solitons. Production of the data set of (a) optical soliton and (c) 1st-order compound derivative, as well as the division of training and test data. (b) Scheme of the SCPINN.

## 3. Results and analysis

In the all-optical soliton communication, the balance of the GVD and SPM in the fiber produces the picosecond optical soliton. The transmission of the femtosecond optical pulse in the fiber needs to consider the influence of the higher-order effects such as TOD, SS and SFS. We use the SCPINN to study the influences of these effects on pulse evolution, and predict the nonlinear dynamics of the picosecond bright and dark solitons, femtosecond two-soliton molecule and three-soliton molecule. These temporal optical solitons have applications in optical switching [39] and optical soliton sources [40]. In particular, soliton molecule have become an ideal carrier of signals in all optical soliton communication because they maintain an optimal distance between solitons, and effectively avoid the occurrence of interaction. In addition, the picosecond bright/dark soliton solutions satisfy the focusing NLSE and the defocusing NLSE degenerated by Eq. (1) respectively, and the femtosecond two-soliton molecule and three-soliton molecule solutions satisfy Eq. (1).

### 3.1. Predicting the formation process of the picosecond bright soliton

In the anomalous dispersion region of the fiber, the effects of GVD and SPM produce the results of negative and positive chirps on the pulse respectively. Therefore, when the chirp of chirped pulse is opposite to the chirp generated by the effects of GVD and SPM, there will appear different phenomenon from that of chirp-free pulse. When the picosecond pulse transmits in the optical fiber, the governing equation is the focused NLSE ignoring the higher-order effects in Eq. (1). The incident pulse is chosen as a hyperbolic secant pulse $Q(0,t) = \sqrt{p_0} \operatorname{sech}(p_0 t) e^{(\frac{-ict^2}{2T_0} + i)}$, where $p_0, T_0, c$ represent the peak power, half width and the chirp coefficient respectively [41]. If the half width $T_0$ and peak power $p_0$ make $L_D / N_{NL} = 1$, the hyperbolic secant pulse incident into the lossless fiber can achieve distortion free transmission, such hyperbolic secant pulse has been used to transmit signals in optical soliton communication [42]. We discuss the cases with the chirp coefficient as $c = 0, c > 0$ for only the effect of GVD and the chirp coefficient as $c = 0, c < 0$ for only the effect of SPM.

The influence of GVD and SPM predicted by the SCPINN on the pulse is shown in Fig. 2. The chirp coefficient $c$ of the incident pulse is an important sensitive parameter of the pulse, the positive and negative chirp affect the compression and broadening in the time /frequency domain during the pulse evolution. When the chirp coefficient $c = 0$, $c > 0$ or $c < 0$ of the incident pulse, it will not affect that the SCPINN accurate characterizes the effect of GVD or SPM on the pulse. The GVD makes the pulse produce the negative chirp, which decreases along the transmission distance. The transmission speed of high-frequency component of the chirp-free pulse is greater than that of low-frequency component, which results in the broadening of the chirp-free pulse (see Figs. 2(a) and 2(b)). Along the transmission distance, the frequency chirp caused by the GVD offsets the initial positive chirp of the positive chirped pulse, which makes the overall negative chirp. Therefore, the positive chirped pulse is first compressed and then widened (see Figs. 2(c) and 2(d)). To show the variation of the spectrum caused by the SPM, we consider the influence of the SPM on the evolution of pulse under the limit condition of zero dispersion. The positive chirp generated by the SPM leads to the broadening of the spectrum of chirp-free pulses and the emergence of symmetrical structures (see Figs. 2(e) and 2(f)). For the pulse with the negative chirp, the chirped effect of the SPM slowly compensates the chirp of the initial pulse, which results in the compressed spectrum of the pulse (see Figs. 2(g) and 2(h)). Moreover, the initial pulse at $z = 0$ in the training set and the output pulse at $z = 5$ in the test set predicted by the SCPINN are almost consistent with the simulation results (see Figs. 2(f) and 2(h)). The convergence theory of PINN [43] shows that the neural network itself has some errors when fitting the data in the training set. This error will increase slowly with the increase of the prediction distance, and it is mainly concentrated in places with large gradients, such as the peak of the spectrum (see Figs. 2(f) and 2(h)).

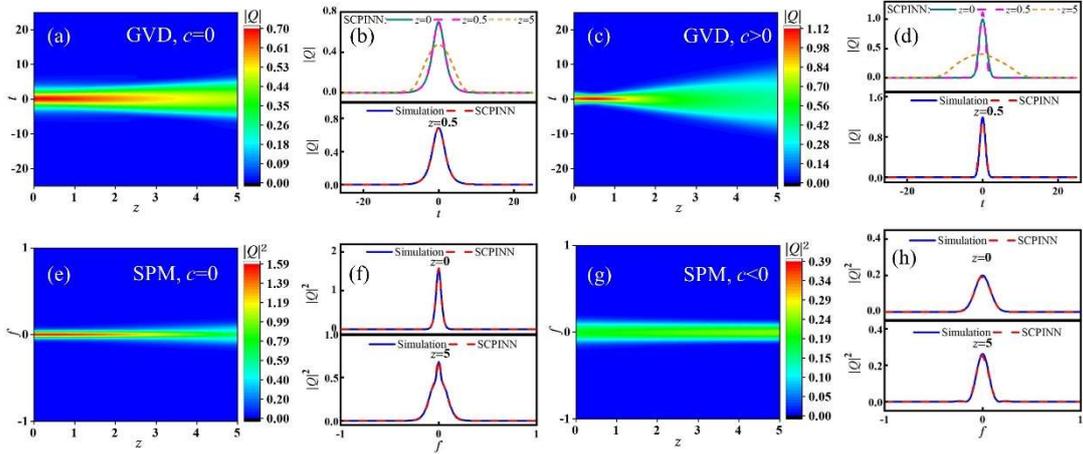

**Fig.2.** Predicting effects of GVD and SPM. (a) The evolution and (e) spectrum diagrams of the chirp-free pulse with $c = 0$. (c) The evolution diagram of the chirped pulse with $c > 0$, and (g) the spectrum diagram of the chirped pulse with $c < 0$. (b) and (d) The variation of the pulse width at distance $z = 0, 0.5, 5$ and the pulse amplitude at distance $z = 0.5$ correspinding to (a) and (c) respectively. (f) and (h) the comparisons of the spectral intensity at propagation distance $z = 0, 5$ correspinding to (e) and (g) respectively.

In the anomalous dispersion region of the optical fiber, when the effects of GVD and SPM are completely balanced, the incident chirp-free pulse maintains the original shape and energy, and turns into a stable bright soliton along the distance. To compares the accuracy of the PINN with SCPINN for predicting bright soliton, the chang of the power $\int_{-\infty}^{\infty} |Q|^2 dt$ for bright soliton along the transmission distance is analyzed by these two neural networks in Fig. 3(a). The power predicted by the PINN decreases sharply along the distance, and soliton is unstable dueing the attenuation process. In contrast, the results predicted by the SCPINN are more accurate. Fig. 3(b) shows the dynamics of the first-order compound derivative predicted by the SCPINN, and the predicted quantity of $\int_{-\infty}^{\infty}|Q_t|^2 dt$ is also relatively accurate. Fig. 3(c) exhibits the comparison of the convergence of the loss function by these two neural networks. The convergence value of loss function for the PINN is about 10 times that of the SCPINN. The curve of the loss function for the SCPINN fluctuates less and the optimization process is more stable. Simultaneously, the insets in Fig. 3(c) reflect that the predicted error of the SCPINN is less than that of the PINN in the whole spatiotemporal region.

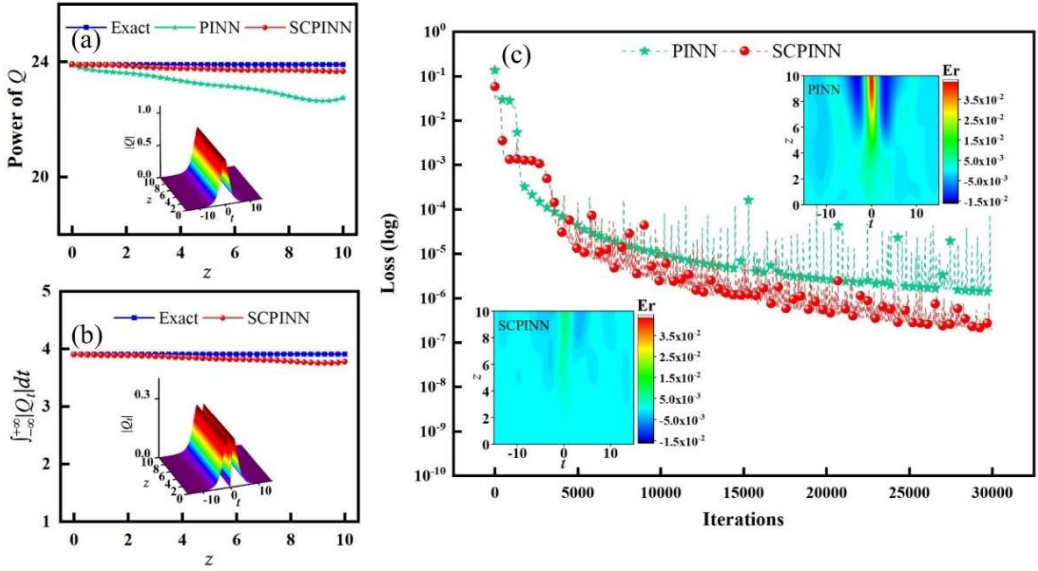

**Fig.3.** Predicting bright soliton. (a)The power and (b) the quantity $\int_{-\infty}^{\infty}|Q_t|^2 dt$ of bright soliton along the propagation distance with the insets respectively showing the dynamics of bright soliton formed by the chirp-free pulse and its 1st-order compound derivative. (c) The convergence curve of the loss function v.s. the iterations for the PINN and SCPINN with the insets showing the error plots predicted by the PINN and SCPINN

### 3.2. Predicting the formation process of the picosecond dark soliton

When the loss coefficient in the fiber is determined, the transmission of dark soliton is more stable than that of bright soliton. Ignoring the higher-order effects in Eq. (1) and considering the defocused NLSE, we study the impacts of GVD and SPM on the dark soliton pulse, whose incident pulse is $Q(0,t) = (B\tanh(B(t-B\sqrt{1-B^2}))-i\sqrt{1-B^2})e^{i\pi}$ with $B$ indicating the depth of the concave part, and it has been used for signal transmission[44] due to its low transmission loss, good anti-interference and low amplified spontaneous incident noise. The dark soliton with $B=1$ is called black soliton.

In Fig. 4, we use the SCPINN to predict variations of the amplitude, phase and power of dark soliton pulse. These predictions are almost consistent with the simulation. In Figs. 4(a)-4(c), for only the effect of GVD on the dark soliton pulse, the power holds constant. The negative chirp generated by the GVD makes the concave part of the dark soliton pulse slightly wider, and two sides of the dark soliton pulse become no longer flat and occur symmetrical oscillations, which become more and more intense along the transmission distance. The reason is that the negative chirp caused by the GVD becomes smaller and smaller, which leads to the broadening of dark soliton pulse. Figs. 4(d)-4(f) show that with the increase of evolution distance, SPM does not change the light intensity of dark soliton pulse, and the light intensity of the concave part for the dark soliton pulse in the center always maintains 0. The phase of the dark soliton pulse changes from time to time, and the phase in the center of the pulse suddenly changes to $\pi$. Compared with Fig. 2, the effect of GVD widens the bright and dark soliton pulses, and the effect of SPM affects the amplitude and spectrum of bright and dark solitons.

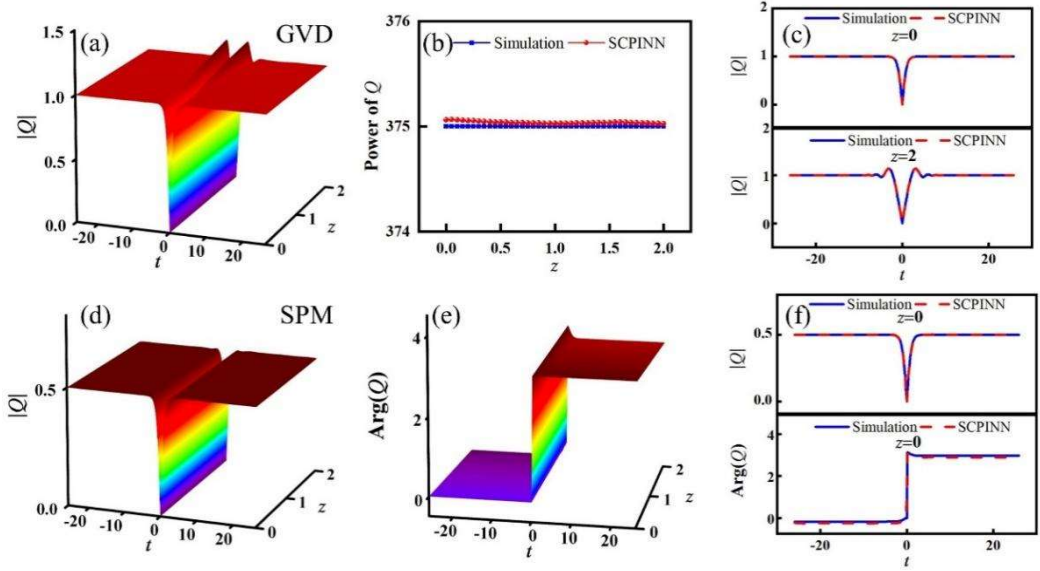

**Fig.4.** Predicting effects of GVD and SPM. (a) The evolution dynamics, (b) power prediction and (c) comparisons of the amplitude of dark soliton pulse at $z = 0, 2$ under the effect of GVD. (d) The evolution dynamics, (e) phase evolution and (f) comparisons of phase and amplitude of dark soliton pulse at $z = 0$ under the effect of SPM.

In the normal dispersion region of the optical fiber, the chirped effect generated by the GVD and SPM counteracts each other, and the transmission of a stable dark soliton is formed in Fig. 5(a). Fig. 5(b) shows the comparison of the accuracy between the PINN and SCPINN for predicting dark soliton. From the evolution of pulse intensity on the uniform background predicted by the PINN at $t = -6$ in Fig. 5(b), the error increases sharply along the transmission distance. However, the predicted results by the SCPINN are almost consistent with the exact solution, and the SCPINN with first-order compound derivative(1st-order) performs best. For the prediction of intensity evolution at the concave position of dark-soliton, the predicted result by the PINN is similar to that by the SCPINN, and there will be no sharp increase of the error.

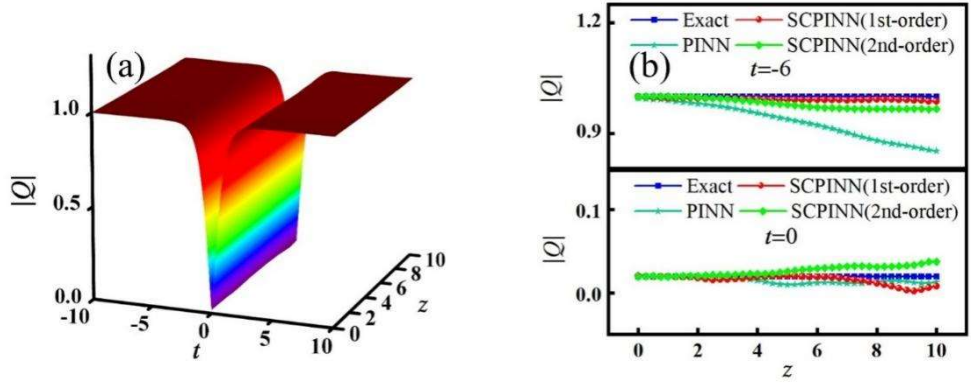

**Fig.5.** Predicting dark soliton. (a) The evolution of soliton and (b) the pulse amplitude at different time predicted by the PINN and SCPINN with 1st-order and 2nd-order compound derivatives.

### 3.3. Predicting the formation process of the femtosecond soliton molecule

As discussed earlier, the various physical effects have the similar impacts on bright and dark solitons [45, 46]. In order to pay attention to the influence of physical effects on the interaction between solitons and the generalization ability of the SCPINN, we study the formation process of soliton molecule under the high-order effects. When the femtosecond optical pulse propagates in optical fiber, the influence of various high-order effects can not be ignored. If these parameters of HNLSE are $\lambda_1 = 0.5, \lambda_2 = 1, \lambda_3 = 0.1, \lambda_4 = 0.6, \lambda_5 = -0.6$, the incident pulses of exact two-soliton molecule and three-soliton molecule [47] are respectively expressed as

$$Q(0,t) = \frac{\frac{50}{13}(e^{-\frac{13}{25}i} + e^{\frac{13}{25}i}) - \frac{100}{29}(e^{-\frac{29}{50}i} + e^{\frac{29}{50}i}) + \frac{40}{11}(e^{-\frac{3t}{100}} - e^{\frac{3t}{100}})(e^{-\frac{11}{20}t} - e^{\frac{11}{20}t})}{\frac{1250}{377}(e^{-\frac{13}{25}t} + e^{\frac{13}{25}t})(e^{-\frac{29}{50}t} + e^{\frac{29}{50}t}) + \frac{400}{121}e^{-\frac{3z}{100}}(e^{-\frac{11}{20}t} - e^{\frac{11}{20}t})^2}. \quad (7)$$

And

$$Q(0,t) = \frac{g_1 + g_3 + g_5}{1 + f_2 + f_4 + f_6}, \quad (8)$$

in which

$g_1 = e^{t(0.7+0.3i)} + e^{6+t(0.8+0.3i)} + e^{12+t(0.9+0.3i)}$,
$g_3 = 0.0441 e^{t(2.2+0.3i)+6} + 9.708 \times 10^{-3} e^{t(2.3+0.3i)+12} + 7.886 \times 10^{-3} e^{t(2.4+0.3i)+18} + 7.53 \times 10^{-3} e^{t(2.5+0.3i)+24} + 1.068 \times 10^{-3} e^{t(2.6+0.3i)+30}$,
$g_5 = 2.13 \times 10^{-10} e^{t(3.9+0.3i)+24} + 5.91 \times 10^{-10} e^{t(4+0.3i)+30} + 1 \times 10^{-10} e^{t(4.1+0.3i)+36}$,
$f_2 = 0.391 e^{t(2.3+0.3i)+12} + 0.444 e^{t(2.2+0.3i)+6} + 0.444 e^{1.5t+6} + 0.510 e^{1.4t} + 0.692 e^{1.7t+18} + 0.781 e^{1.6t+12} + 0.309 e^{1.8t+24}$,
$f_4 = 3.937 \times 10^{-6} e^{3t+12} + 2.452 \times 10^{-5} e^{3.1t+18} + 4.313 \times 10^{-5} e^{3.2t+24} + 1.483 \times 10^{-5} e^{3.3t+30} + 1.444 \times 10^{-6} e^{3.4t+36}$,
$f_6 = 3.552 \times 10^{-15} e^{4.8t+36}$.

When the pulse wavelength is near the zero dispersion wavelength of the fiber, the TOD must be considered during the process of evolution. As shown in Figs 6 (a) and 6 (c), the effect of TOD alone does not affects the power change in the process of soliton transmission, namely $\int_{-\infty}^{\infty} |Q|^2 dt = \text{constant}$ and $\int_{-\infty}^{\infty} |Q_t|^2 dt = \text{constant}$. With the increase of transmission distance $z$, an oscillating

pulse is formed near the front edge of pulse and the pulse becomes steeper. Under the influence of the TOD, the two soliton pulses weaken their amplitudes, split into the multiple pulses and disperse their energy. Comparing Figs. 6(b) and 6(d), the error of the prediction via the SCPINN is mainly concentrated in the complicated oscillation region of the evolution plots for the derivative term, whose reason is that the derivatives of the higher-order differential terms have bad rule and their training is not easy.

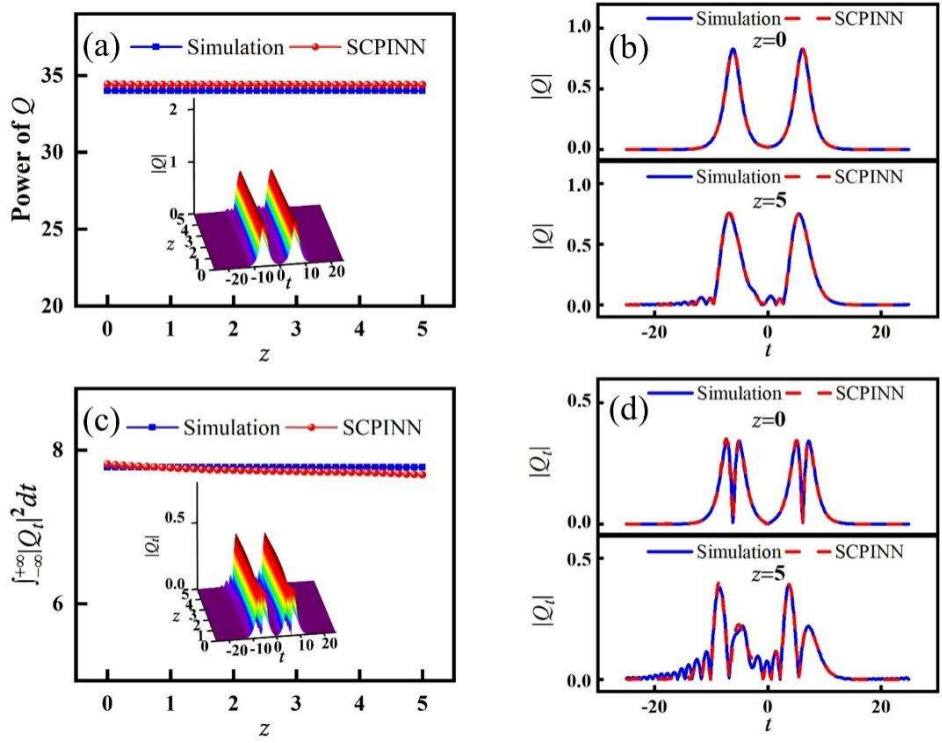

**Fig.6.** Predicting effect of TOD. (a) The power and (c) the quantity $\int_{-\infty}^{\infty}|Q_t|^2 dt$ of the two-soliton molecule along the propagation distance with the insets showing the dynamics of two-soliton molecule and the 1st-order compound derivative predicted by the SCPINN. (b) and (d) The comparison of amplitudes of two-soliton molecule and the related first-order compound derivative at distance $z = 0, 5$.

We find that the shift of two-soliton molecule pulse happen under both SS and SFS effects. The SS makes the two-soliton molecule pulse shift towards the backward edge (see Figs. 7(a) and 7(d)), and yet the SFS causes two-soliton molecule pulse to shift towards the front edge (see Figs. 7(b) and 7(d)). When the effects of SS and SFS work together, the direction of shift of two-soliton molecule pulse is consistent with that of the SS alone (see Figs. 7(c) and 7(d)). Under the same intensity of the SS or SFS, the influence of the SS on the femtosecond pulse is greater than that of the SFS. Under the three physical scenarios, the energy of the output pulse remains unchanged, that is $\int_{-\infty}^{\infty}|Q|^2 dt = \text{constant}$. The higher-order effects do not produce the gain or loss on the two-soliton molecule pulse (see Fig. 7(e)). Compared with the PINN, the SCPINN is more accurate in predicting the higher-order physical effects and has less relative error $L_2$ (see Fig.7(f)).

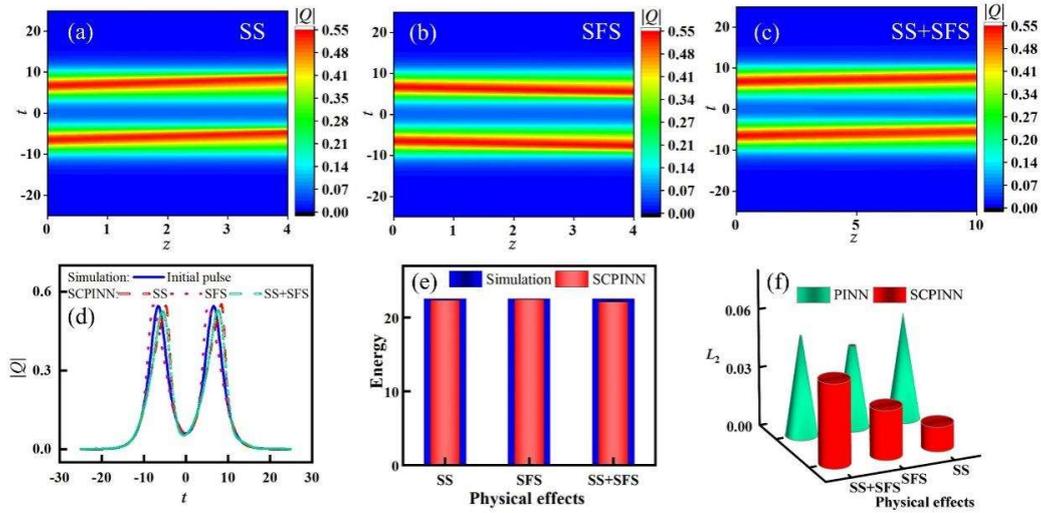

**Fig.7.** Predicting effects of SS and SFS. The evolution of two-soliton molecule under (a) the SS, (b) the SFS, and (c) SS and SFS and the corresponding (d) amplitude and (e) energy of the output pulse. (f) Relative error $L_2$ of soliton molecule predicted by the PINN and SCPINN under different physical effects.

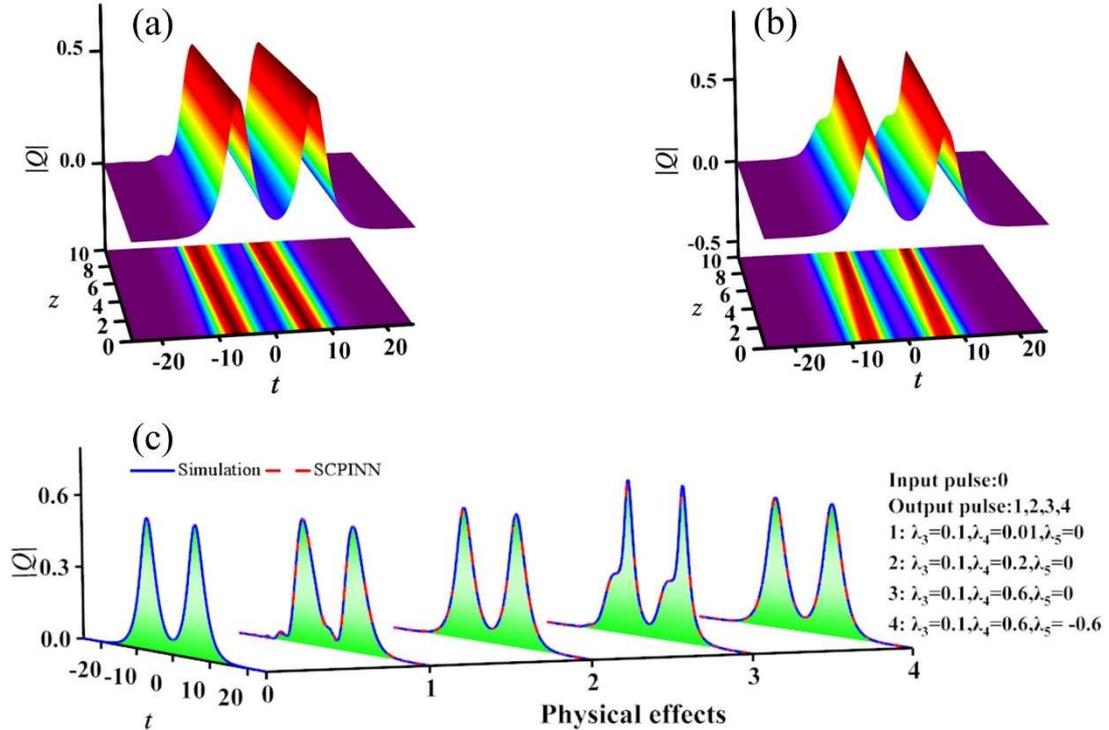

**Fig.8.** Predicting effects of TOD, SS and SFS. The evolution of two-soliton molecule with parameters (a) $\lambda_3 = 0.1, \lambda_4 = 0.01, \lambda_5 = 0$ and (b) $\lambda_3 = 0.1, \lambda_4 = 0.6, \lambda_5 = 0$. (c) The output pulse at $z = 10$ for different intensities of the higher-order effects with the same input pulse.

For the Hirota equation, the effect of SS balances the effects of TOD and SFS, and solitons can maintain their original shapes during the transmission (see Fig. 8(c)). Next, we consider the relationship between the effects of TOD and SS. When the intensities of TOD and SS are $\lambda_3 = 0.1$, $\lambda_4 = 0.01$ respectively, an oscillating pulse is formed near the pulse front edge and the pulse becomes steeper which is simlilar to the characteristics of the TOD. When the intensity of the SS increases to $\lambda_4 = 0.2$, the effects of SS and TOD are almost completely offseted, and the pulse transmission of soliton molecule is basically not affected by these two higher-order effects (see Figs. 8(a) and 8(c)), that is, there is a suitable combination of physical parameters of the TOD and SS, which makes two effects completely offset and realizes stable soliton transmission. If the intensity of the SS effect continues to increase, the pulse moves backward edge and the peak increases, and the transmission of two-soliton molecule shows the similar characteristics of the SS effect (see Fig. 8 (b) and 8 (c)), which shows that the influence of the SS is greater than that of the TOD in Hirota equation. In fact, when the femtosecond optical pulse propagates in the optical fiber, the effects of GVD and SPM, and the higher-order effects balance each other to form a stable transmission of soliton. In Fig. 8, the pulse transmission predicted by the SCPINN is almost consistent with the simulation results.

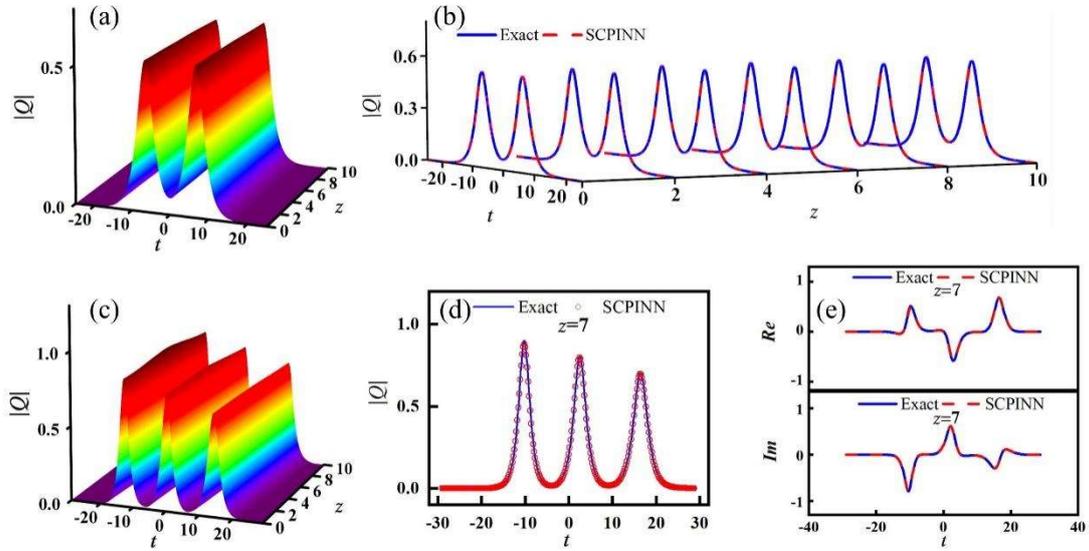

**Fig.9.** Predicting soliton molecules. The pulse evolutions of (a) two-soliton molecule and (c) three-soliton molecule. (b) Waterfall comparison diagram of evolution of two-soliton molecule along the distance. Comparison of (d) pulse amplitudes and (e) the real and imaginary parts of three-soliton molecule at $z = 7$.

The pulse compression caused by the nonlinear effect and the pulse broadening caused by the dispersion effect are balanced, and thus the femtosecond optical soliton pulse forms in the optical fiber and transmits stably. Figs. 9(a) and 9(c) show the pulse evolution diagram of two-soliton molecule and three-soliton molecule predicted by the SCPINN. When the wave velocities between two or three solitons reach a special resonance state, the attraction and repulsion between two or three solitons realize a balance to form a bound state-soliton molecule. Along the transmission distance, the pulse predicted by the SCPINN decays slowly. In the optical fiber, the soliton molecule can usually be obtained by the splitting of single pulse or balance between multiple pulses. Fig. 9(b) shows the two-soliton molecule predicted by the SCPINN. Each soliton maintains its own shape during the transmission, and this characteristic can greatly reduce the bit error rate in the optical fiber communication. Figs. 9(d) and 9(e) exhibit the comparison between the predicted and exact pulses at $z=7$. Whether the real part, imaginary part or amplitude of the pulse predicted by the SCPINN are consistent with the fitting of the exact solution.

In addition, Table 1 shows that with the number of neural network layers increases, the SCPINN more and more accurately reconstructs the two-soliton molecule in Fig. 9(a), but the process of improving the accuracy is accompanied by an increase in time consumption. When the layers of SCPINN exceed 5, the neural network may have over fitting. The excessive weights and biases cause the neural network to be difficult to train and reduce its prediction ability, and 5 layers are exactly the best choice for the SCPINN.

Table 1. The relative error $L_2$ and time cost of two-soliton molecule predicted by the SCPINN with different layers with 40 neurons per layer

| Layer | 1 | 2 | 3 | 4 | 5 | 6 | 7 |
|---|---|---|---|---|---|---|---|
| $L_2$ | 31.54% | 21.98% | 12.21% | 6.05% | 0.58% | 6.62% | 4.03% |
| Time | 1528.5s | 4885.1s | 8192.6s | 11465.3s | 14911.3% | 18498.5s | 21838.3s |

4. **Discussion and analysis**

Table 2 shows the hyperparameters of the two neural networks. The difference is that the SCPINN is a parallel two-subnets design and introduces a flexible learning rate. Under the same data set, the accuracy of two networks in predicting soliton dynamics is tested.

Table 2. Hyperparameters of the PINN and SCPINN

|  | Network structure | Optimizer | Iterations | Learning rate |
|---|---|---|---|---|
| PINN | 5 layers*40 neurons | Adam | 30000 | 0.0001 |
| SCPINN | 2subnets*5layers*40 neurons | Adam | 30000 | $0.001 \times 0.9^{\text{int}(\frac{\text{Current iterations}}{1000})}$ |

Figs. 10(a)-10(d) display the comparison of the convergence of loss function for the different soliton dynamics predicted by the PINN and SCPINN with 1st-order and 2nd-order compound derivatives. The results show that the convergence value of loss function of the SCPINN with the compound derivative information is smaller, and the curve of loss function is smoother, namely the probability falling into the local optimization in the process of neural network optimization is lower. Fig. 10(e) clearly shows that the relative error $L_2$ of the SCPINN in predicting soliton dynamics is smaller than that of the PINN, in which the SCPINN(1st-order) performs best, and the accuracy is about 5-11 times that of the PINN. With the increasing complexity of soliton dynamics, the relative error $L_2$ of neural network prediction will gradually add. To measure the increasing rate of relative error $L_2$ for different neural networks, we translate the curve of the relative error $L_2$ of PINN over that of the SCPINN. In Fig. 10(e), the pink dotted line is the translation line of the PINN relative error $L_2$. It can be clearly found that the curve slope of relative error $L_2$ of the SCPINN(1st-order) is obviously smallest, that is, the error of the SCPINN(1st-order) increases most slowly. Because the 2nd-order compound derivative $r_{tt}, m_{tt}$ has more complicated bad rule, such as unsmooth curve, etc., it is more difficult to optimize and train the SCPINN (2nd-order) in Figs. 10(a)-10(d).

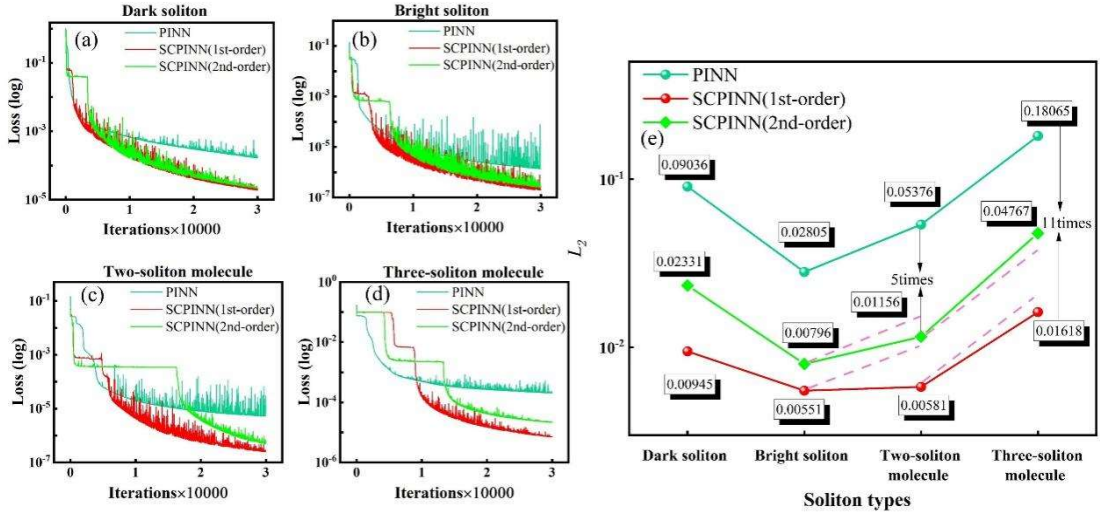

**Fig.10.** The convergence diagram of loss function of (a) dark soliton, (b) bright soliton, (c) two-soliton molecule and (d) three-soliton molecule predicted by the PINN and SCPINN; (e) Relative error $L_2$ of the PINN and SCPINN when predicting different soliton dynamics.

## 5. Conclusion

To sum up, we propose a SCPINN with the parallel two-subnets to the formation process of bright and dark picosecond optical solitons, and femtosecond soliton molecule, and predict the nonlinear dynamics of these solitons in single-mode fiber. A variety of physical quantities such as the power conservation, pulse amplitude and phase change are systematically analyzed and verified. In addition, from the perspective of the convergence of loss function and the accuracy of network prediction, we comprehensively analyze and compare the PINN and SCPINN. The results show that SCPINN is basically consistent with the reference solution in characterizing the pulse evolution under the physical effects of GVD, SPM, TOD, SS and SFS, as well as the dynamics of soliton transmission in the fiber. Moreover, compared with PINN, the prediction accuracy of the SCPINN for soliton pulse evolution is improved by an order of magnitude, the maximum relative error is $L_2 \approx 1.62\%$. Therefore, as a calculation method combining deep learning and physical laws, the SCPINN is not only an effective solver of PDEs, but also a forward-looking method to study the modeling and analysis of soliton dynamics in the fiber.

**Acknowledgements**

Zhejiang Provincial Natural Science Foundation of China (Grant No. LR20A050001); National Natural Science Foundation of China (Grant Nos. 11874324, 12075210, 12275240). Scientific Research and Developed Fund of Zhejiang A&F University (Grant No. 2021FR0009).

**Conflict of interest**

The authors have declared that no conflict of interest exists.

**Ethical Standards**

This Research does not involve Human Participants and/or Animals.

**Appendix A. Some of the codes**

This Appendix accompanies the main manuscript, and contains critical codes of all algorithms in the paper that aim to provide a reference for researchers. Data underlying the results presented in this paper are not publicly available at this time but may be obtained from the authors upon reasonable request.

The complex neural network $Q(z,t) = (u(z,t), v(z,t))$ can be defined as

```
def net_uv(self, z, t):
    X = tf.concat([z,t],1)
    uv = self.neural_net_sin(X, self.weights, self.biases, self.ahs)
    u = uv[:,0:1]
    v = uv[:,1:2]
    return u,v
```

PINN：

Moreover, the PINN f(z,t) is chosen the form

```
def net_f_uv(self, z, t):
```

```
u, v = self.net_uv(z,t)
u_t = tf.gradients(u, t)[0]
v_t = tf.gradients(v, t)[0]
u_z = tf.gradients(u, z)[0]
u_tt = tf.gradients(u_t, t)[0]
u_ttt = tf.gradients(u_tt, t)[0]
v_z = tf.gradients(v, z)[0]
v_tt = tf.gradients(v_t, t)[0]
v_ttt = tf.gradients(v_tt, t)[0]
f_u = u_z + 0.5 * v_tt + (u ** 2 + v ** 2) * v + 0.1 * u_ttt + 0.6 * (u ** 2 + v ** 2) * u_x
f_v = v_z - 0.5 * u_tt - (u ** 2 + v ** 2) * u + 0.1 * v_ttt + 0.6 * (u ** 2 + v ** 2) * v_x
return f_u, f_v
```

SCPINN(1st-order):

The compound derivative neural network $Q_t(z,t) = (u1(z,t), v1(z,t))$ can be defined as

```
def net_uv_ compound derivative (self, z, t):
    X = tf.concat([z, t], 1)
    u1v1 = self.neural_net_sin(X, self.weights1, self.biases1, self.ahs1)
    u1 = u1v1[:, 0:1]
    v1 = u1v1[:, 1:2]
    return u1, v1
```

Moreover, the SCPINN(1st-order) f(z,t) is chosen the form

```
def net_f_uv(self, z, t):
    u1, v1= self.net_uv_ compound derivative (z, t)
    u, v, = self.net_uv(z, t)
    u_t = tf.gradients(u, t)[0]
    v_t = tf.gradients(v, t)[0]
    u_z = tf.gradients(u, z)[0]
    u_tt = tf.gradients(u_t, t)[0]
    u_ttt = tf.gradients(u_tt, t)[0]
    v_z = tf.gradients(v, z)[0]
    v_tt = tf.gradients(v_t, t)[0]
    v_ttt = tf.gradients(v_tt, t)[0]
    f_u = u_z+ 0.5 * v_tt + (u ** 2 + v ** 2) * v + 0. * u_ttt + 0.6 * (u ** 2 + v ** 2) * u_t
    f_v = v_z - 0.5 * u_tt - (u ** 2 + v ** 2) * u + 0.1 * v_ttt + 0.6 * (u ** 2 + v ** 2) * v_t
    f_u1 = u1 - u_t
```

```
        f_v1 = v1- v_t
        return f_u, f_v, f_u1, f_v1
```

SCPINN(2st-order):

The compound derivative neural network $Q_{tt}(z,t) = (u2(z,t), v2(z,t))$ can be defined as

```
def net_uv_ compound derivative (self, z, t):
        X = tf.concat([z, t], 1)
        u2v2 = self.neural_net_sin(X, self.weights1, self.biases1, self.ahs1)
        u2 = u2v2[:, 0:1]
        v2 = u2v2[:, 1:2]
        return u2, v2
```

Moreover, the SCPINN(2st-order) f(z,t) is chosen the form

```
def net_f_uv(self, z, t):
        u2, v2= self.net_uv_ compound derivative (z, t)
        u, v, = self.net_uv(z, t)
        u_t = tf.gradients(u, t)[0]
        v_t = tf.gradients(v, t)[0]
        u_z = tf.gradients(u, z)[0]
        u_tt = tf.gradients(u_t, t)[0]
        u_ttt = tf.gradients(u_tt, t)[0]
        v_z = tf.gradients(v, z)[0]
        v_tt = tf.gradients(v_t, t)[0]
        v_ttt = tf.gradients(v_tt, t)[0]
        f_u = u_z + 0.5 * v_tt + (u ** 2 + v ** 2) * v + 0. * u_ttt + 0.6 * (u ** 2 + v ** 2) * u_t
        f_v = v_z - 0.5 * u_tt - (u ** 2 + v ** 2) * u + 0.1 * v_ttt + 0.6 * (u ** 2 + v ** 2) * v_t
        f_u2 = u2 - u_tt
        f_v2 = v2- v_tt
        return f_u, f_v, f_u2, f_v2
```